\documentclass[aps,prl,reprint,superscriptaddress]{revtex4-1}
\usepackage{graphicx}
\usepackage{color}

\begin{document}


\title{Whirling spin order in the quasicrystal approximant Au$_{72}$Al$_{14}$Tb$_{14}$}


\author{Taku J Sato}
\email[]{taku@tagen.tohoku.ac.jp}
\affiliation{Institute of Multidisciplinary Research for Advanced Materials, Tohoku University,
  2-1-1 Katahira, Aoba, Sendai 980-8577, Japan}

\author{Asuka Ishikawa}
\author{Akira Sakurai}
\author{Masashi Hattori}
\affiliation{Department of Materials Science, Tokyo University of Science,
  Katsushika, Japan}

\author{Maxim Avdeev}
\affiliation{Australian Centre for Neutron Scattering, Australian Nuclear Science and Technology Organisation,
  Locked Bag 2001, Kirrawee, NSW 2232, Australia}
\affiliation{School of Chemistry, The University of Sydney, Sydney, NSW 2006, Australia}

\author{Ryuji Tamura}
\affiliation{Department of Materials Science, Tokyo University of Science,
  Katsushika, Japan}


\date{\today}

\begin{abstract}
  Neutron powder diffraction experiment has been performed on the quasicrystal approximant Au$_{72}$Al$_{14}$Tb$_{14}$, a body-center-cubic crystal of icosahedral spin clusters.
  The long-range antiferromagnetic order was confirmed at the transition temperature $T_{\rm N} = 10.4$~K.
  The magnetic structure consists of noncoplanar whirling spins on the icosahedral clusters, arranging antiferroic-manner.
  A simple icosahedral spin-cluster model with uniaxial anisotropy accounts well the whirling spin order as well as the in-field metamagnetic transition, indicating that the icosahedral symmetry is essential.
\end{abstract}


\maketitle

Magnetic clusters with icosahedral point symmetry have attracted continuous interest because of nontrivial ground and/or excited states originating from frustrated geometry~\cite{Coffey1992,Axenovich2001,Schnack2001,Schmidt2003,Schnack2004,Luban2004,SchroderC05,Konstantinidis2011,Konstantinidis2015}.
Prominent examples may be a noncoplanar order with large skyrmion number for classical Heisenberg spins in a truncated icosahedron~\cite{Coffey1992}, and a sequence of rotational bands formed as low-lying excitations for quantum Heisenberg spins in an icosidodecahedron~\cite{Schnack2001}.
The icosahedral symmetry is incompatible with lattice translation, and hence they usually do not form interacting periodic array.
Instead, they form either quasiperiodic array, as in magnetic quasicrystals~\cite{Goldman2014,Sato2005}, or periodic array with negligible inter-cluster interactions, as in the Keplerate molecular magnets~\cite{Muller2001}.
Recently, it has been recognized that crystalline phases exist nearby the quasicrystalline phase in phase diagram that have icosahedral clusters arranging periodically.
They are now called ``quasicrystal approximants''.
The approximants offer a new playground to study the magnetic behavior of interacting icosahedral clusters in a periodic lattice.

Magnetic approximants have been found in various rare-earth ($R$) based alloy systems, for instance Cd$_6R$ ~\cite{Gomez2003,Tsai2013}, Ag-In-$R$~\cite{Morita2008,Ibuka2011}, Au-Al-$R$~\cite{Ishikawa2016,Nakayama2015}, and Au-Si-$R$~\cite{Hiroto2013,Gebresenbut2014}, to note a few.
Most of those magnetic approximants show either ferromagnetic or spin-glass-like behavior at low temperatures, except for binary Cd$_6$$R$.
The former ferromagnetic order is rather trivial, where the symmetry of the cluster becomes less effective for magnetism.
The latter glassy behavior may be related to geometrical frustration expected for the icosahedral symmetry clusters.
However, the glassy freezing, possibly due to disorder inevitable in real alloy systems, conceals their intrinsic nature.
In either case, the magnetic diffraction has only limited ability in determining microscopic magnetic structures since magnetic Bragg peaks superimposedly appear on strong nuclear Bragg peaks in the former case~\cite{Gebresenbut2014}, or only diffuse scattering appears in the latter~\cite{Ibuka2011}.
Hence, the antiferromagnetic order has been sought after in approximants.
Up to now, the Cd$_6$$R$ compounds are the only approximants that show clear antiferromagnetic long-range order, evidenced by the bulk magnetization~\cite{Tamura2010}, neutron diffraction~\cite{Ryan2019}, and x-ray resonant scattering~\cite{Kim2012}.
However, their magnetic structures have never been solved, partly because their structural phase transitions introduce complicated domain formations~\cite{Tamura2002}, and also because strong absorption due to Cd for neutrons or heavier elements for x-rays makes quantitative magnetic structure analysis quite difficult.
Hence, microscopic understanding of magnetic ordering in the approximants has been far from accomplished.

\begin{figure}
  \includegraphics[scale=0.35, angle=90, trim= 100 100 100 100 ]{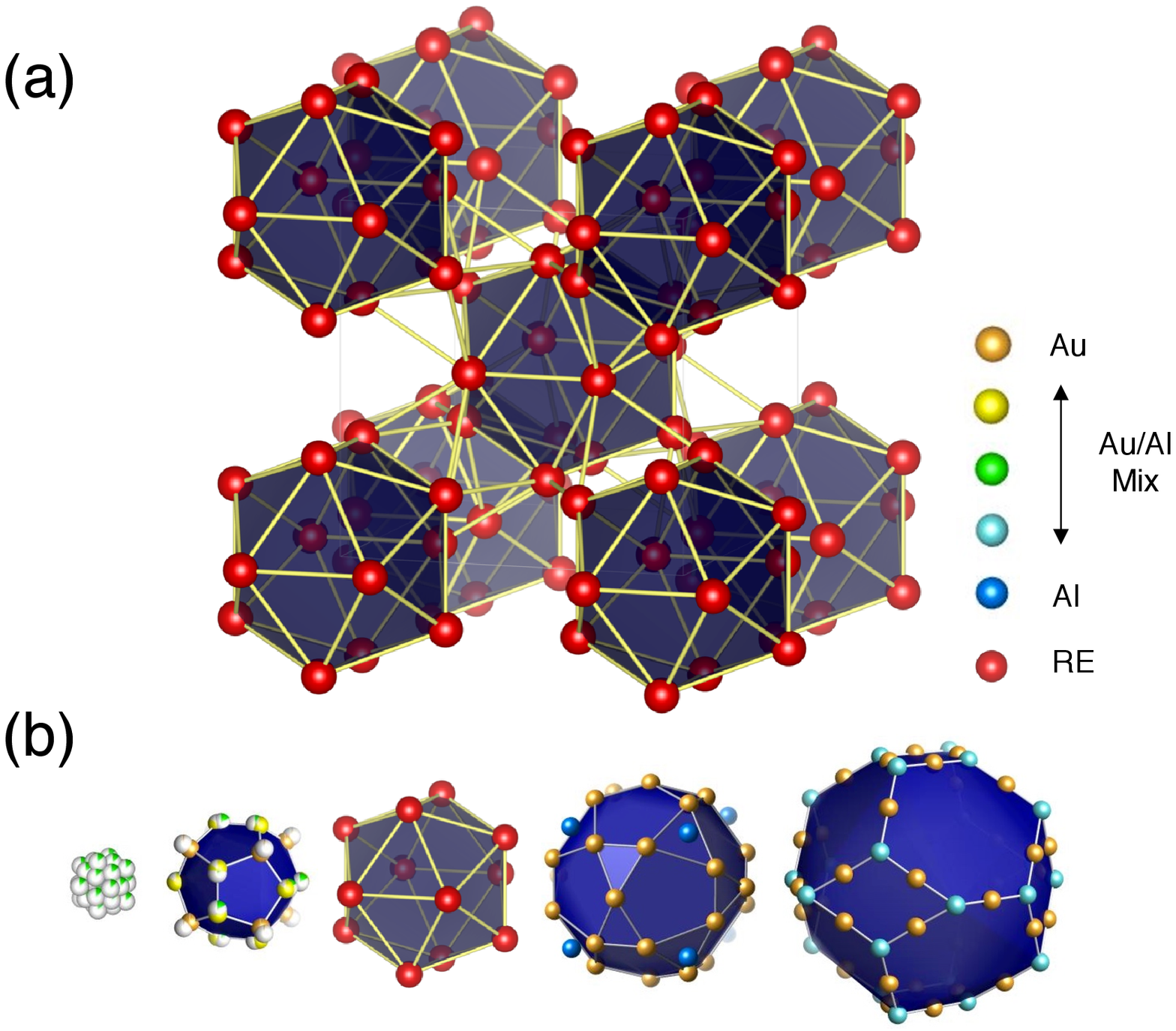}
  \caption{\label{figure1} (a) Body-center-cubic array of Tsai-type icosahedral clusters in the Au-Al-Tb approximant.
    (b) Multiple shell structure of the Tsai-type cluster.
    Magnetic Tb$^{3+}$ ions occupy the second shell, selectively.
  }
\end{figure}

Recently, a new magnetic approximant was found in the Au$_x$Al$_{86-x}R_{14}$ system~\cite{Ishikawa2016}.
This approximant has body-centered cubic (bcc) structure with the space group $Im\bar{3}$ [Fig.~\ref{figure1}(a)].
It consists of multiple shell clusters of slightly-distorted icosahedral symmetry, known as the Tsai-type clusters [Fig.~1(b)].
The rare-earth atoms selectively occupy the second icosahedral shell, and hence the system can be regarded as the bcc array of icosahedral spin clusters.
For $R =$ Tb, the approximant phase forms in a wide composition range of $49 < x < 72$, and shows various magnetic ground states ranging from antiferromagnetic to spin-glass-like orders depending on $x$~\cite{Ishikawa2018}.
Specifically at $x = 72$, the magnetic susceptibility shows Curie-Weiss behavior at high temperatures with the effective moment $\mu_{\rm eff} = 9.85\mu_{\rm B}$.
The Weiss temperature is estimated as $\Theta_{\rm p} = 4.2$~K, suggesting dominant ferromagnetic interaction.
At $T_{\rm N} = 11.8$~K, a clear anomaly was detected in the magnetic susceptibility; no irreversibility between zero-field-cooling and field-cooling runs was observed below $T_{\rm N}$, ruling out the possibility of the spin-glass-like freezing.
The decreasing magnetic susceptibility at lower temperature, instead, indicates the antiferromagnetic long-range order~\cite{Ishikawa2018}.

As the Au-Al-Tb approximant does not include the strong neutron absorber, such as Cd or Gd, it becomes possible to study microscopic magnetic order of the periodically arrayed icosahedral spin clusters using this compound.
Hence,  in the present work we have performed the neutron powder diffraction.
We found that the magnetic structure is far from a simple N\'{e}el order, but is interesting noncoplanar whirling spin order in the icosahedral clusters.
The spins at the opposite vertices of the cluster align antiparallelly, and hence the total magnetic moment of each cluster is exactly zero.
The magnetic order can also be interpreted as antiferroic arrangement of cluster magnetic-toroidal multipoles, breaking the bcc centering-translation invariance.
A simple model spin-Hamiltonian for single icosahedral cluster is proposed, reproducing observed magnetic structure as well as the bulk magnetization behavior.

A polycrystalline alloy of Au$_{72}$Al$_{14}$Tb$_{14}$ was prepared by arc melting with high purity ($> 99.9$~wt\%) Au, Al and Tb elements~\cite{Ishikawa2018}.
The as-cast alloy was annealed at 1073~K for 50~h under an inert Ar atmosphere, and then quenched into chilled water to obtain a single phased sample.
The phase purity of the sample was examined using the powder x-ray diffraction (Rigaku MiniFlex600) and the scanning electron microscopy (JEOL JSM-IT100).

The neutron powder diffraction experiment has been performed using the high-resolution powder diffractometer ECHIDNA installed at the OPAL reactor, Australian Nuclear Science and Technology Organisation~\cite{Avdeev2018}.
For the magnetic diffraction measurements, neutrons with $\lambda = 2.4395$~\AA\ were selected using the Ge 331 reflections, whereas for the structure analysis, to obtain reflections in a wide $Q$-range, we select $\lambda = 1.622$~\AA\ using the Ge 335 reflections.
The sample was set in the double cylindrical annular can made of vanadium to reduce the absorption effect of Au.
The sample was set to the cold head of the closed cycle $^4$He refrigerator with the base temperature 3.5~K.
Obtained powder diffraction patterns were analyzed using the Rietveld method combined with magnetic representations analysis, performed using the home-made magnetic structure analysis code~\cite{MSAS2019}.


\begin{figure*}
  \includegraphics[scale=0.36, angle=-90]{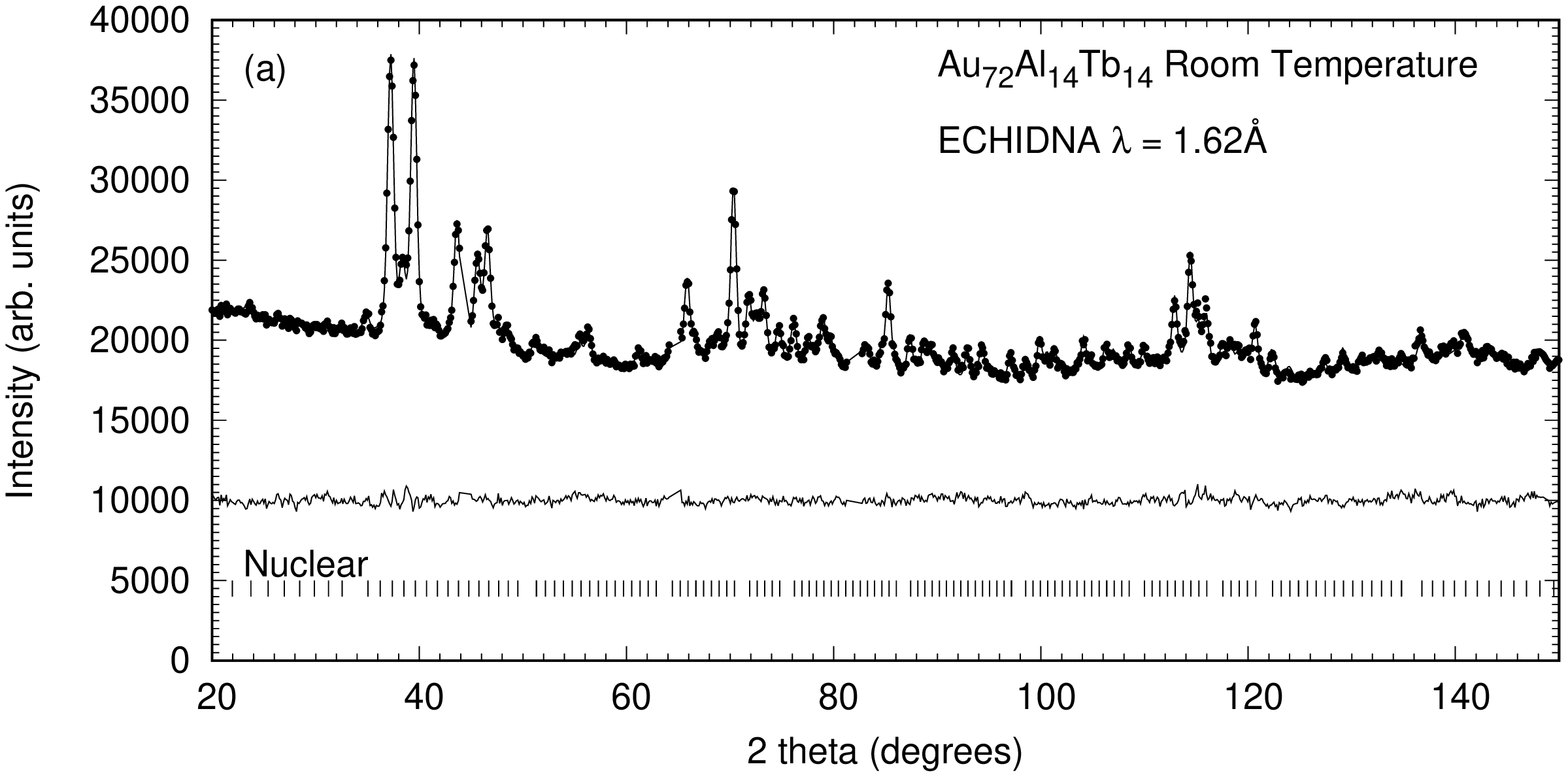} 
  \includegraphics[scale=0.26, angle=-90]{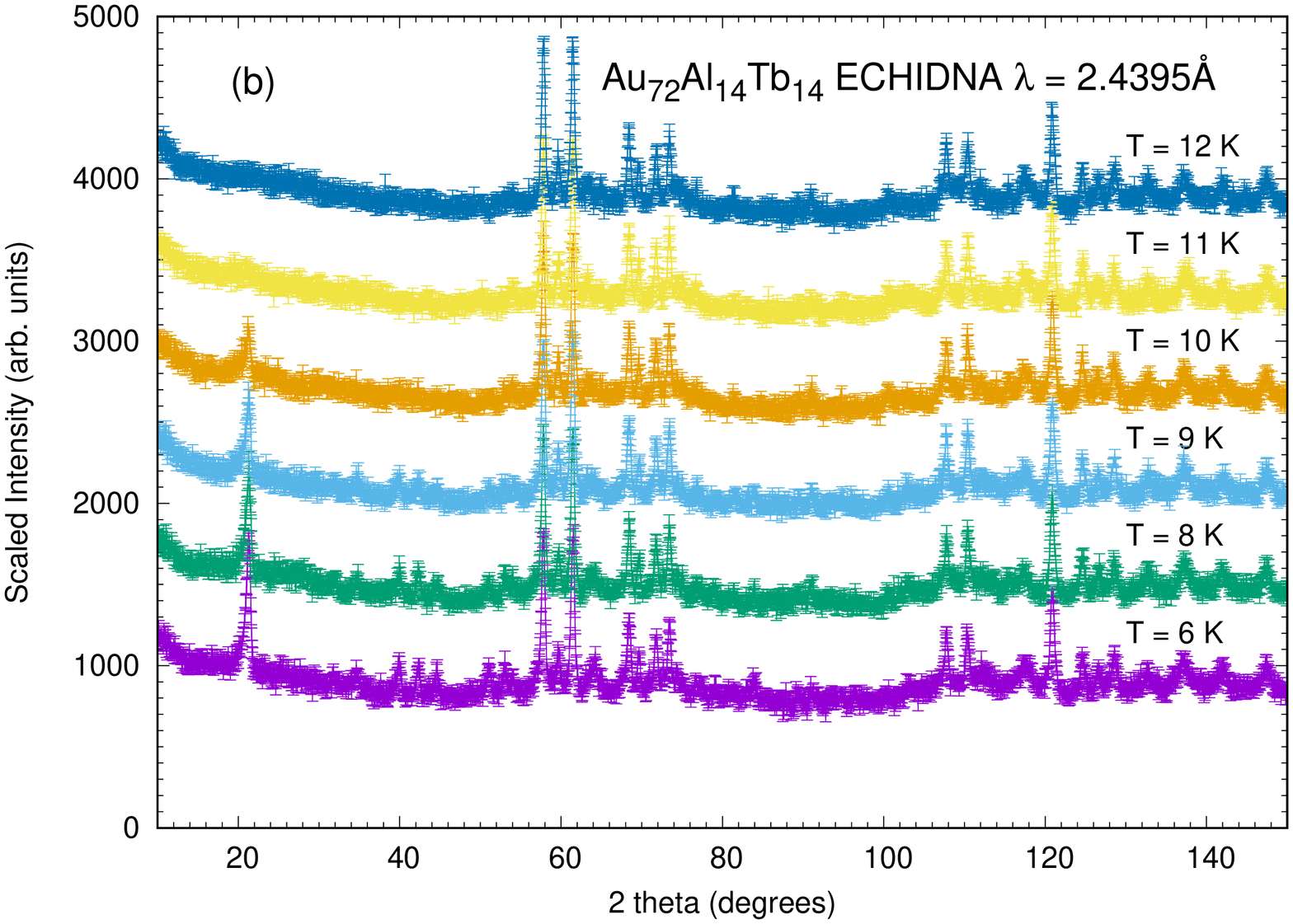}
  \includegraphics[scale=0.36, angle=-90]{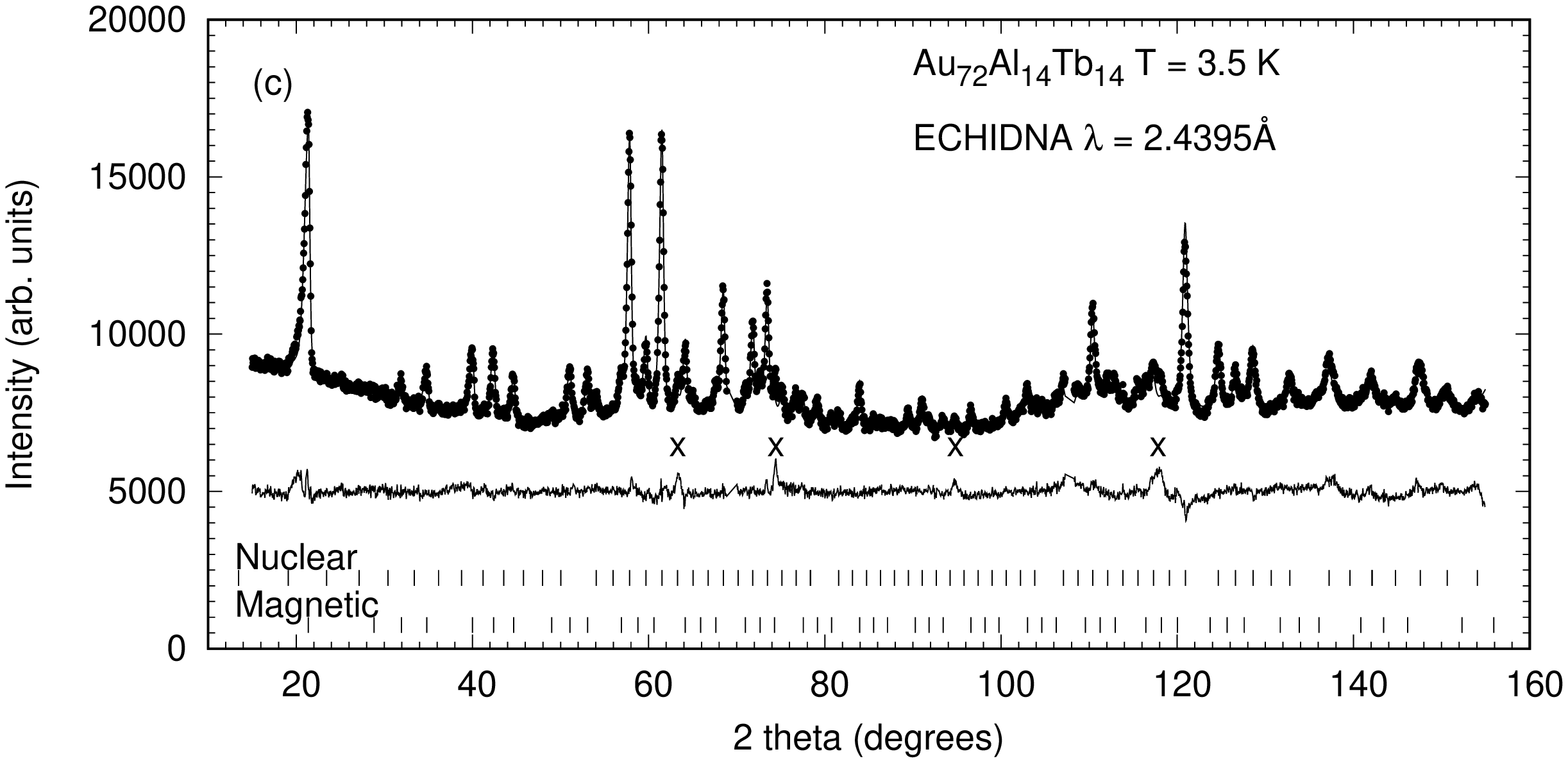}
  \includegraphics[scale=0.26, angle=-90]{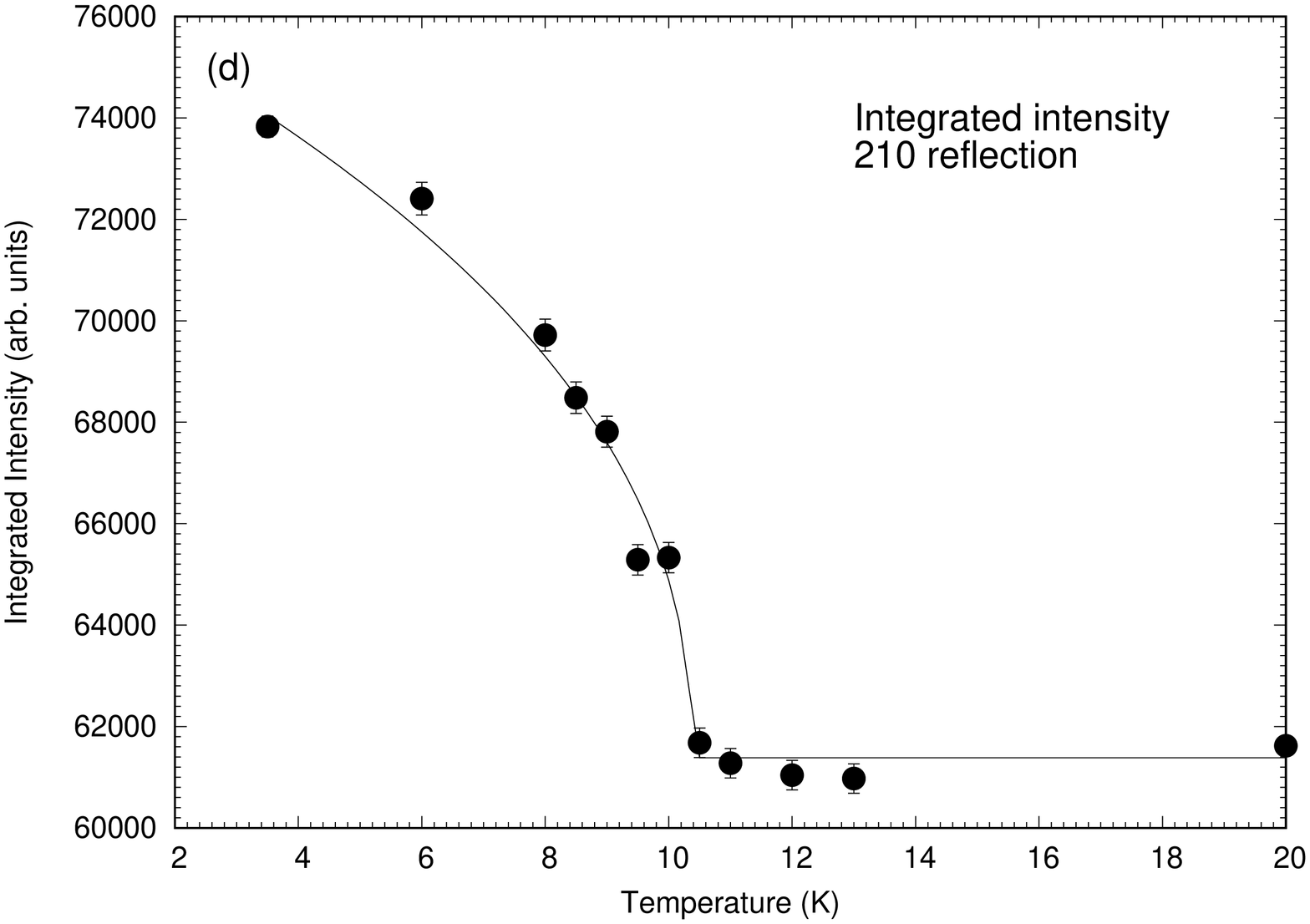}
  \caption{\label{figure2}(a) Neutron powder diffraction pattern at the room temperature.
    (b) Low-temperature powder diffraction patterns at $T = 12, 11, 10, 9, 8$ and 6~K (from top to bottom).
    (c) Neutron powder diffraction pattern at $T = 3.5$~K.
    (d) Temperature dependence of the integrated intensity of the 210 magnetic reflection.
    In (a) and (c), the nuclear and magnetic reflection positions are denoted by the vertical lines at the bottom.
    The Rietveld fitting results are shown by the solid lines, whereas the difference between the observation and the fitting is also shown below the observation/fitting result.
    $2\theta$-regions where the Bragg peaks from the vanadium sample can appear were removed from the fitting.
    The ``x'' marks in the bottom panel indicate nuclear reflections from unknown impurity phase.
  }
\end{figure*}

First, the structure refinement was performed to confirm the consistency with the earlier report~\cite{Ishikawa2018}.
Figure~\ref{figure2}(a) shows the resulting neutron powder diffraction pattern measured at the room temperature.
The pattern was analyzed using the Rietveld method with the reported crystallographic parameters as the initial parameters.
The refined parameters are given in Table~SI (Supplemental), whereas the fitting result, as well as difference from the observation, is shown in Fig.~\ref{figure2}(a).
The refined parameters are in good agreement with the single-crystal x-ray results, confirming the high quality of the powder sample.

\begin{figure}
  \includegraphics[scale=0.35,angle=90, trim=00 00 0 50]{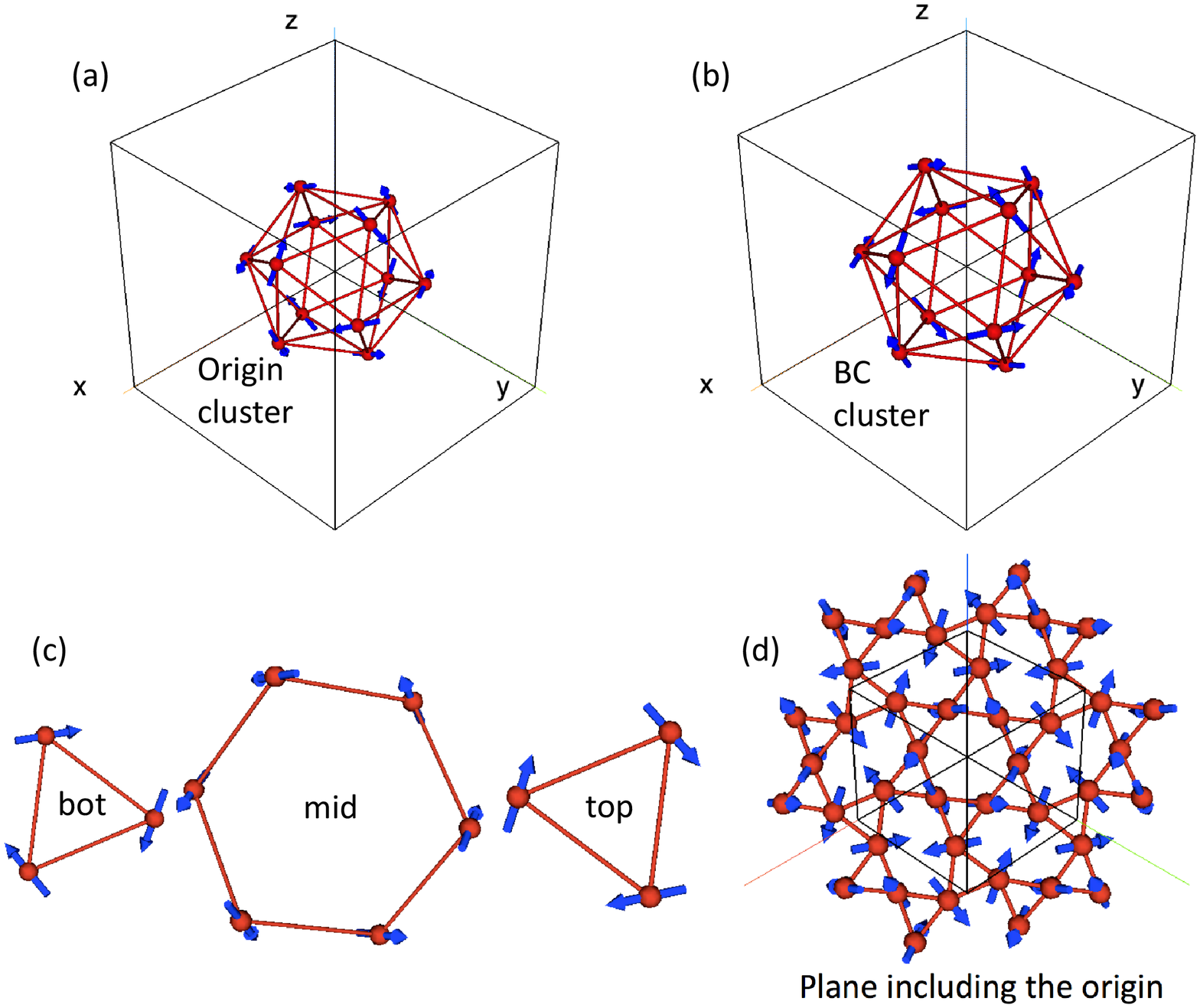} 
  \caption{\label{figure3}(a) and (b) Obtained spin configurations in icosahedral spin clusters at (a) the origin and at (b) the body center.
    The clusters are depicted along the [111] direction.
    (c) Layer by layer illustration of spin ordering in the center cluster shown in (a). 
    (d) A slice of spin structure in the plane perpendicular to the [111] axis through the origin.
  }
\end{figure}

Neutron powder diffraction patterns at the low temperatures $T = 3.5, 6, 8, 9, 10, 11$ and 12~K are shown in Fig.~\ref{figure2}(b) and \ref{figure2}(c).
As the temperature is lowered, new Bragg reflections appear in the low-$2\theta$ region, being a clear indication of magnetic long-range order. 
The magnetic reflections appear at the forbidden reflection positions of the bcc lattice, {\it i.e.} $hkl$ with $h + k + l = 2n + 1$ ($n$: integer).
This indicates that corresponding magnetic order is antiferromagnetic, as suggested from the bulk magnetic measurement, and breaks the body-center symmetry of the underlying crystalline lattice.

Temperature dependence of integrated intensity for the magnetic 210 reflection is shown in Fig.~\ref{figure2}(d).
By fitting the temperature dependence to the power law $I \propto (T_{\rm N} - T)^{2\beta}$ for $T < T_{\rm N}$, we estimate the antiferromagnetic transition temperature as $T_{\rm N} = 10.4(3)$~K.
This is in reasonable agreement with the transition temperature (11.8~K) obtained in the bulk magnetic measurement; slight difference may be due to the temperature calibration in the powder diffraction experiment and/or difficulty in obtaining precise transition temperature using bulk susceptibility data.

The magnetic structure is obtained using the diffraction pattern at the base temperature $T = 3.5$~K [Fig.~\ref{figure2}(c)] with the aid of the magnetic representation analysis.
The details of the analysis are given in Supplemental information.
A combination of two basis vectors (BVs) in the single irreducible representation ($\nu = 2$) reproduces the observed magnetic diffraction pattern satisfactorily.
The coefficients for the two BVs are $C_1^2 = 6.71(5)$ and $C_2^2 = -3.43(7)$, which give rise to the magnetic moment size of $7.5(3)~\mu_{\rm B}$ for the Tb$^{3+}$ ions at $T = 3.5$~K.
The calculated Rietveld profile is shown in Fig.~2(c) to be compared with the observation.
It may be noted that the ordered moment is rather small compared to the free Tb$^{3+}$ moment size $g_{J} \mu_{\rm B} J = 9~\mu_{\rm B}$, where $J = 6$ is the total angular momentum and $g_{J} = 3/2$ is the Land\'{e} $g$-factor for Tb$^{3+}$.
Indeed, the 210 reflection intensity still increases at $T = 3.5$~K, the lowest temperature achievable in the present setup, suggesting it would increase to 9~$\mu_{\rm B}$ for $T \rightarrow 0$.

The obtained magnetic structure is schematically shown in Fig.~3.
The magnetic structure has quite a few characteristics that are rarely seen in ordinary antiferromagnets.
First, the magnetic order comprises noncoplanar whirling arrangement of Tb spins.
This can be seen the best by depicting spin configurations of icosahedral clusters at the origin and body center separately, as shown in Figs. 3(a) and 3(b).
The spins are in the mirror plane of the icosahedral cluster, and are almost tangential to the cluster surface, resulting in the whirling configuration around the [111] axis.
(Angle between the spin vector $\vec{J}_i$ and its position vector from the origin $\vec{r}_i$ is $\simeq 86^{\circ}$.)
The spin directions of corresponding Tb sites of the origin and body-center clusters are antiparallel, indicating that bcc symmetry is broken by the ``antiferroic'' arrangement of cluster spins.
Note that the Tb spins at the opposite vertices of the single cluster are also antiparallel, and hence the total magnetic moment of a single cluster is exactly zero.
By applying cluster multipole description~\cite{Suzuki2017}, we found that the third order ($p = 3$) magnetic-toroidal multipole remains finite for each cluster, changing its sign from origin to the body center (Supplemental information).
Therefore, we can regard the long-range order as antiferroic order of cluster magnetic-toroidal multipoles, breaking the bcc translational symmetry. 

\begin{figure}
  \includegraphics[scale=0.32, angle=-90]{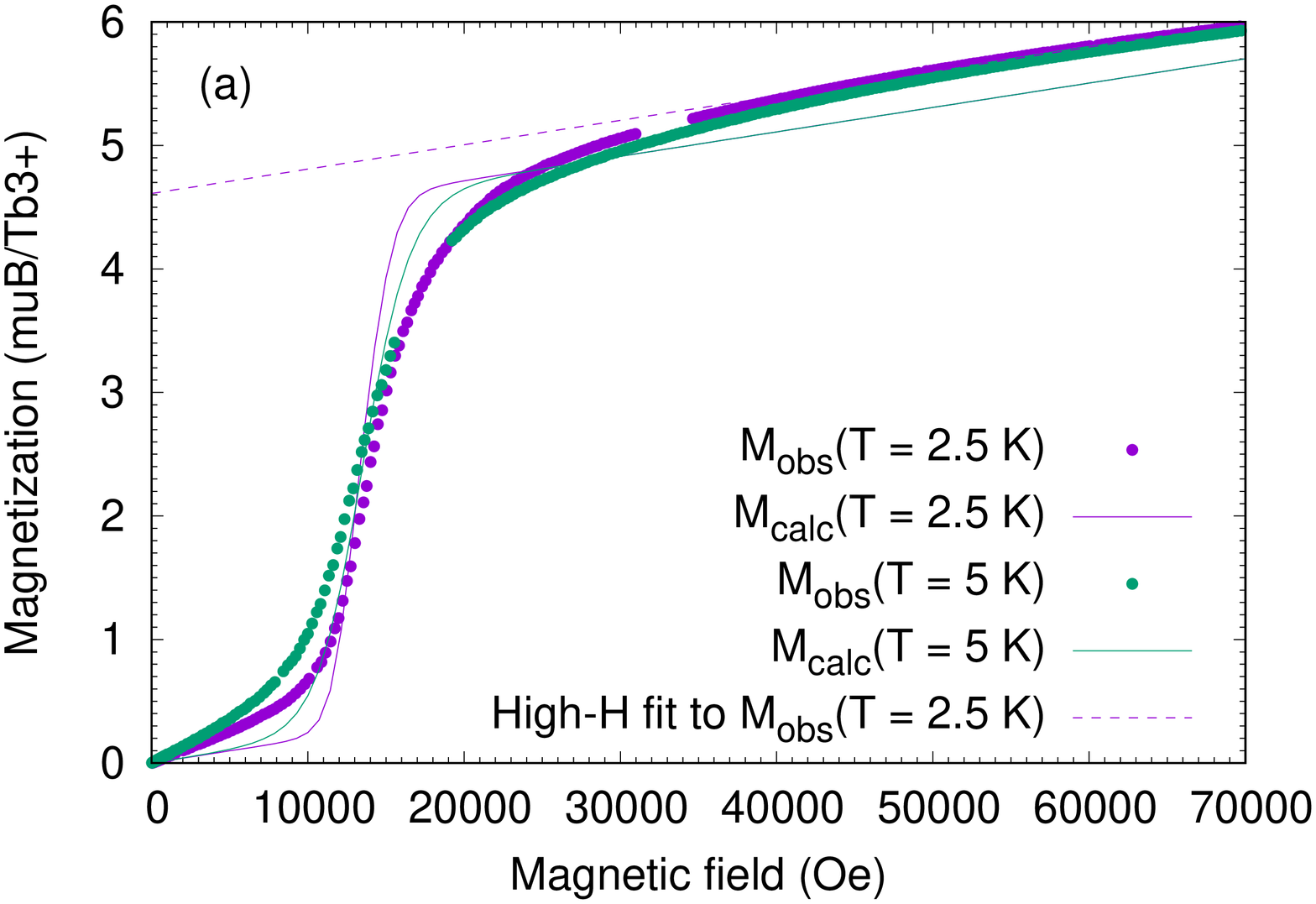} 
  \includegraphics[scale=0.45, angle=-90, trim = 100 250 100 250]{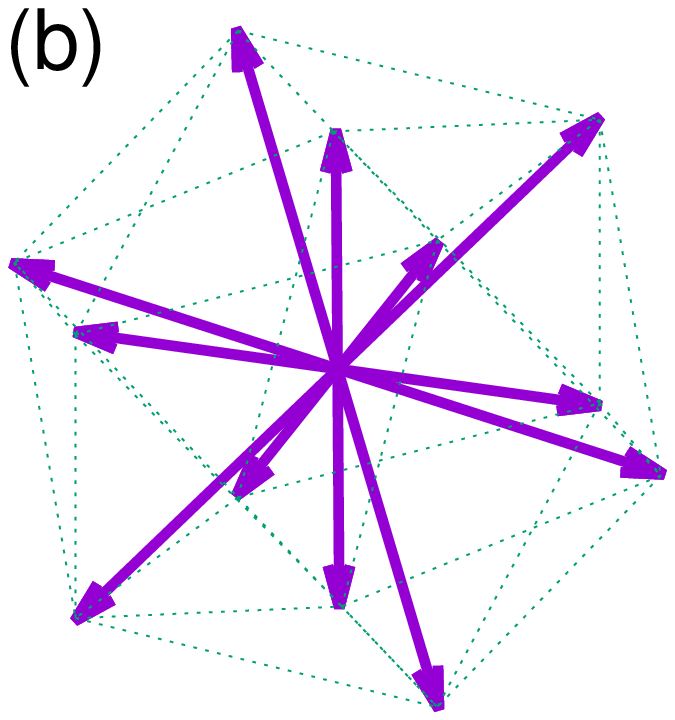}
  \includegraphics[scale=0.45, angle=-90, trim = 100 250 100 250]{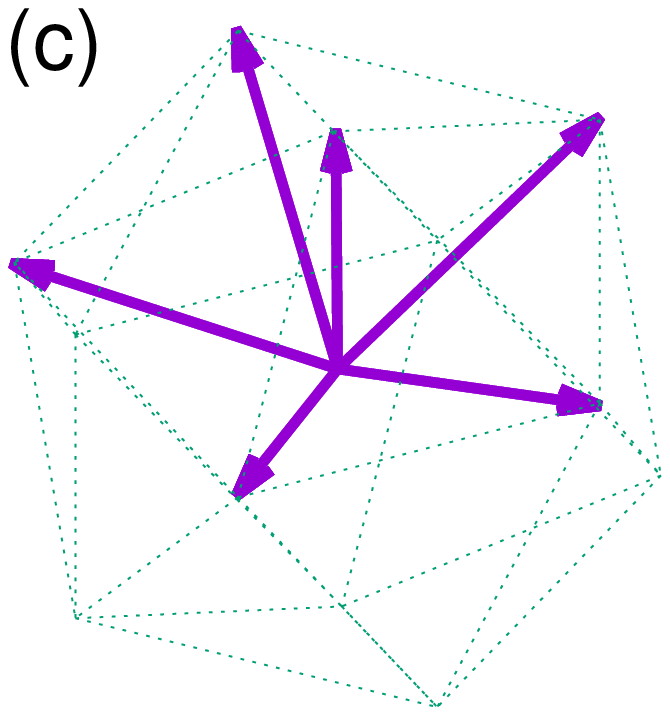}
  \caption{\label{figure4}(a) Powder averaged magnetization curves calculated at $T = 2.5$ (magenta) and 5~K (green) using the single icosahedral spin cluster model introduced in the main text, whereas corresponding dots stand for the experimental results reported in Ref.~\onlinecite{Ishikawa2018}.
    Magenta dashed line stands for the high field ($H > 40000$~Oe) fit to estimate van Vleck contribution.
    (b) and (c) Simulated spin configurations in one icosahedral cluster in the low field [(b); $H < 13000$~Oe] and high field [(c); $H > 13000$~Oe].
  }
\end{figure}

The 12 spin vectors in one icosahedron point to vertices of icosahedron, as shown in Fig.~\ref{figure4}(b).
This, at first glance, looks similar to the situation of magnetic skyrmions~\cite{MuhlbauerS09}, where the spin rotation can be characterized by the continuous vector field spreading $4\pi$.
However, in the present case the chirality of spin configuration is different from layer to layer; as seen in Fig.~\ref{figure3}(c); the Tb spins in the top and bottom triangle layers exhibit the same clockwise rotation, whereas the opposite rotation in the mid buckling hexagon layer.
Hence, the skyrmion number cannot be defined, in contrast to the theoretical prediction for the C60-type magnetic cluster~\cite{Coffey1992}.

As described above, the obtained magnetic structure shows a number of intriguing characteristics awaiting further microscopic understanding of its origin.
Here, we propose a very simple model that reproduces the observed structure as a first approximation.
The ordered moments are along high symmetry direction, which is the surface tangential direction in a local mirror plane.
This strongly suggests that the moment direction is fixed by the icosahedral symmetry, and further suggests easy-axis anisotropy along the ordered moment direction by the crystalline electric field (CEF).
The metamagnetic transition takes place at the finite magnetic field ($H \simeq 13.6$~kOe) at low temperatures~\cite{Ishikawa2018}, further supporting the existence of the easy-axis anisotropy which confines the Tb$^{3+}$ moment to its two-fold (Ising-type) CEF ground state.
Based on the above observations, we introduce the simplest model Hamiltonian for the Au-Al-Tb approximant as follows:
\begin{equation}
  {\cal H} = -\sum_{<i,j>} J_{ij} \vec{J}_i \cdot \vec{J}_j - g_J \mu_{\rm B} \sum_{i} \vec{J}_i \cdot \vec{H}_{\rm ext},
\end{equation}
where $J_{ij}$ stands for the exchange interaction between the $i$-th and $j$-th sites, $\vec{H}_{\rm ext}$ stands for the external magnetic field, and $\vec{J}_i$ represents the Ising spin vectors of Tb$^{3+}$, and is restricted either parallel or antiparallel to the ordered moment direction, {\it i.e.}, $\vec{J}_i = \pm 6 \vec{n}_i$, where $\vec{n}_i$ is the unit vector along the local anisotropy direction.
We only take account of nearest-neighbor ($J_1$) and next-nearest neighbor ($J_2$) pairs in a single isolated icosahedral spin cluster as a minimal model, {\it i.e.}, $1 \leq i,j \leq 12$.

Classical energy of all the possible $2^{12}$ spin configurations was numerically calculated for various combination of spin interactions.
As a result, we found that the observed magnetic structure is stabilized as a ground state for dominant ferromagnetic next-nearest neighbor $J_2$.
The ground state is uniquely selected, and hence strictly speaking there is no indication of geometrical frustration for the present case.
Under a certain magnetic field, the flipping of half of spins takes place from the configurations shown in Fig.~\ref{figure4}(b) to \ref{figure4}(c).
This results in a net magnetic moment in a single icosahedral cluster, reproducing the metamagnetic transition observed in the bulk magnetization.
Addition of weak nearest neighbor ferromagnetic interaction $J_1 > 0$ does not alter the ground state as long as $ J_1 < J_2/2$.
It, however, certainly reduces the critical field for the metamagnetic transition, as the higher-energy state becomes stable under magnetic field.
By tuning $J_2$ as well as the ratio $J_1/J_2$, we found that both the Weiss temperature and metamagnetic transition field are well reproduced with $J_2 = 0.6$~K and $J_1 = 0.44 J_2$.
The Weiss temperature for the above parameters is $\Theta = 4.3$~K, whereas the powder averaged metamagnetic transition field is $H_{\rm mt} \simeq 13$~kOe, being in good agreement with the reported values.
The external field dependence of the magnetization per Tb$^{3+}$ spin is simulated as shown in Fig.~4(a).
By adding experimentally estimated van Vleck contribution (linear term), originating from mixing of ground-state and higher-energy CEF wavefunctions, we found that the numerical simulation well reproduces the external field dependence of the magnetization in low-temperature range.
The saturated magnetic moment estimated from the simple icosahedral model is 4.3$\mu_{\rm B}$ per one Tb$^{3+}$ ion, again in good agreement with the experimental estimation, 4.6$\mu_{\rm B}$, obtained by linearly extrapolating the high-field data to $H \rightarrow 0$ shown in Fig.~4(a). 
These results strongly suggest that the simplest model indeed captures essential characteristics of the magnetism of the Au-Al-Tb quasicrystal approximant.
On the other hand, long-range antiferroic order of the cluster (magnetic toroidal) multipoles is definitely due to inter-cluster interactions, and has to be elucidated using more elaborated theory/simulation.

As noted above, with the dominant ferromagnetic next-nearest neighbor interactions (or similarly with the dominant antiferromagnetic nearest-neighbor interactions), the ground state of the single icosahedral cluster is uniquely selected [Figs.~3(a) and 3(b)], and hence non-degenerate.
On the other hand, for the ferromagnetic nearest neighbor and/or antiferromagnetic next-nearest neighbor interactions, we found that the ground states of the icosahedral cluster are highly degenerated because of the competition of inter-spin interactions and local easy-axis anisotropy.
This is exactly the same situation as the ``spin ice'' pyrochlore antiferromagnets, where the local easy-axis anisotropy and ferromagnetic inter-spin interactions result in the intriguing ``ice rule'' degeneracy~\cite{Gingras2011}.

In reality, the icosahedral clusters form a bcc cubic crystal, and thus are not isolated.
The inter-cluster network of Tb spins may be highlighted by the slices perpendicular to the [111] axis through the origin as illustrated in Figs 3(d).
The Tb sites form disordered kagome network in this plane, an archetypal frustrated geometry, although in the present case the Tb sites are not exactly on the plane, but buckling.
Since the three dimensional Tb network can be indeed regarded as the inter-penetrating kagome planes, this compound may be another new highly-frustrated 3D magnet when ferromagnetic nearest-neighbor and/or antiferromagnetic next-nearest neighbor interactions are realized by tuning the composition.
Further study in this direction must be apparently interesting.

In summary, we have elucidated the magnetic structure in the Au$_{72}$Al$_{14}$Tb$_{14}$ quasicrystal approximant using the powder neutron diffraction.
The obtained magnetic structure is the whirling spin order in the icosahedral clusters, with the counter-rotating whirls in the adjacent layers along the crystallographic [111] axis.
The obtained noncoplanar magnetic structure is found to be a result of strong uniaxial anisotropy, together with the dominant next-nearest-neighbor ferromagnetic interactions.

\begin{acknowledgments}

The authors thank T. Sugimoto, K. Morita and T. Hiroto for valuable discussions, and A. P. Tsai for continuous supports.
This work is partly supported by Grants-in-Aids for Scientific Research (15H05883, 16H04018, 17K18744) from MEXT of Japan, and by the research program “Dynamic alliance for open innovation bridging human, environment, and materials.”
The travel expenses for the neutron scattering experiment at ANSTO were partly supported by the General User Program for Neutron Scattering Experiments, Institute for Solid State Physics, University of Tokyo.

\end{acknowledgments}

\bibliography{aualtb}

\end{document}